\documentclass[article,aip,apl,
amsmath,amssymb,
reprint,%
]{revtex4-1}
\usepackage{booktabs} % 用于三线表
\usepackage{multirow}
\usepackage{graphicx}% Include figure files
\usepackage{dcolumn}% Align table columns on decimal point
\usepackage{bm}% bold math
\usepackage{subfigure}
\usepackage{subcaption}
\usepackage{color}
\usepackage{float}
\usepackage[utf8]{inputenc}
\usepackage[T1]{fontenc}
\usepackage{mathptmx}
\usepackage{etoolbox}
\usepackage{threeparttable}
\usepackage[marginal]{footmisc}
\renewcommand{\thefootnote}{}

\makeatletter
\def\@email#1#2{%
 \endgroup
 \patchcmd{\titleblock@produce}
  {\frontmatter@RRAPformat}
  {\frontmatter@RRAPformat{\produce@RRAP{*#1\href{mailto:#2}{#2}}}\frontmatter@RRAPformat}
  {}{}
}%
\makeatother

\begin{document}

\preprint{AIP/123-QED}
\title{Joint gravity survey using an absolute atom gravimeter and relative gravimeters}

%\title{Experimental Study on the Performance of A-Grav Atom Gravimeter in Field Mobile Observations}%Force line breaks with \\
\author{Li Chen-yang$^{\#}$}
\affiliation{Hefei National Research Center for Physical Sciences at the Microscale and School of Physical Sciences, University of Science and Technology of China, Hefei 230026, China}
\affiliation{Shanghai Research Center for Quantum Science and CAS Center for Excellence in Quantum Information and Quantum Physics, University of Science and Technology of China, Shanghai 201315, China}
\affiliation{Hefei National Laboratory, University of Science and Technology of China, Hefei 230088, China}
% \altaffiliation{Physics Department, XYZ University.}%Lines break automatically or can be forced with \\
\author{Xu Ru-gang$^{\#}$}
% \email{Second.Author@institution.edu.}
\affiliation{Anhui Earthquake Agency, Hefei  230031, China}
\affiliation{School of Earth and Space Sciences, University of Science and Technology of China, Hefei   230026, China}
\affiliation{Mengcheng National Geophysical Observatory, University of Science and Technology of China, Mengcheng  233500, China}
\affiliation{Wuhan Gravitation and Solid Earth Tides, National Observation and Research Station, Wuhan  430071, China}

\author{Chen Xi}
\affiliation{Hefei National Research Center for Physical Sciences at the Microscale and School of Physical Sciences, University of Science and Technology of China, Hefei 230026, China}
\affiliation{Shanghai Research Center for Quantum Science and CAS Center for Excellence in Quantum Information and Quantum Physics, University of Science and Technology of China, Shanghai 201315, China}
\affiliation{Hefei National Laboratory, University of Science and Technology of China, Hefei 230088, China}

\author{Sun Hong-bo}
\affiliation{Anhui Earthquake Agency, Hefei  230031, China}

\author{Li Su-peng}
\affiliation{Hefei National Research Center for Physical Sciences at the Microscale and School of Physical Sciences, University of Science and Technology of China, Hefei 230026, China}
\affiliation{Shanghai Research Center for Quantum Science and CAS Center for Excellence in Quantum Information and Quantum Physics, University of Science and Technology of China, Shanghai 201315, China}
\affiliation{Hefei National Laboratory, University of Science and Technology of China, Hefei 230088, China}

\author{Luo Yu}
\affiliation{Shanghai Research Center for Quantum Science and CAS Center for Excellence in Quantum Information and Quantum Physics, University of Science and Technology of China, Shanghai 201315, China}

\author{Huang Ming-qi}
\affiliation{Shanghai Research Center for Quantum Science and CAS Center for Excellence in Quantum Information and Quantum Physics, University of Science and Technology of China, Shanghai 201315, China}

\author{Di Xue-feng}
\affiliation{Hefei National Research Center for Physical Sciences at the Microscale and School of Physical Sciences, University of Science and Technology of China, Hefei 230026, China}
\affiliation{Shanghai Research Center for Quantum Science and CAS Center for Excellence in Quantum Information and Quantum Physics, University of Science and Technology of China, Shanghai 201315, China}
\affiliation{Hefei National Laboratory, University of Science and Technology of China, Hefei 230088, China}

\author{Li Zhao-long}
\affiliation{Hefei National Research Center for Physical Sciences at the Microscale and School of Physical Sciences, University of Science and Technology of China, Hefei 230026, China}
\affiliation{Shanghai Research Center for Quantum Science and CAS Center for Excellence in Quantum Information and Quantum Physics, University of Science and Technology of China, Shanghai 201315, China}
\affiliation{Hefei National Laboratory, University of Science and Technology of China, Hefei 230088, China}

\author{Xiao Wei-peng}
\affiliation{Anhui Earthquake Agency, Hefei  230031, China}

\author{Liang Xiao}
\affiliation{Anhui Earthquake Agency, Hefei  230031, China}

\author{Yang Xuan}
\affiliation{Hefei National Laboratory, University of Science and Technology of China, Hefei 230088, China}

\author{Huang Xian-liang}
\affiliation{Anhui Earthquake Agency, Hefei  230031, China}
\affiliation{Mengcheng National Geophysical Observatory, University of Science and Technology of China, Mengcheng  233500, China}

\author{Yao Hua-jian}
\affiliation{Hefei National Laboratory, University of Science and Technology of China, Hefei 230088, China}
\affiliation{School of Earth and Space Sciences, University of Science and Technology of China, Hefei   230026, China}
\affiliation{Mengcheng National Geophysical Observatory, University of Science and Technology of China, Mengcheng  233500, China}

\author{Huang Jin-shui}
\affiliation{Hefei National Laboratory, University of Science and Technology of China, Hefei 230088, China}
\affiliation{School of Earth and Space Sciences, University of Science and Technology of China, Hefei   230026, China}
\affiliation{Mengcheng National Geophysical Observatory, University of Science and Technology of China, Mengcheng  233500, China}

\author{Chen Luo-kan\thanks{Co-corresponding author: lkchen@ustc.edu.cn}}
 \email{lkchen@ustc.edu.cn}
\affiliation{Hefei National Research Center for Physical Sciences at the Microscale and School of Physical Sciences, University of Science and Technology of China, Hefei 230026, China}
\affiliation{Shanghai Research Center for Quantum Science and CAS Center for Excellence in Quantum Information and Quantum Physics, University of Science and Technology of China, Shanghai 201315, China}
\affiliation{Hefei National Laboratory, University of Science and Technology of China, Hefei 230088, China}

\author{Chen Shuai\thanks{Co-corresponding author: shuai@ustc.edu.cn}}
 \email{shuai@ustc.edu.cn}
% \homepage{https://quantum.ustc.edu.cn/web/index.php/node/85}
\affiliation{Hefei National Research Center for Physical Sciences at the Microscale and School of Physical Sciences, University of Science and Technology of China, Hefei 230026, China}
\affiliation{Shanghai Research Center for Quantum Science and CAS Center for Excellence in Quantum Information and Quantum Physics, University of Science and Technology of China, Shanghai 201315, China}
\affiliation{Hefei National Laboratory, University of Science and Technology of China, Hefei 230088, China}

%\footnotetext[9]{Co-corresponding authors: email1@example.com (Author C), email2@example.com (Author D).}

\date{\today}% It is always \today, today,
             %  but any date may be explicitly specified
%\footnotetext[$\dagger$]{These authors contributed equally to this work.}

%\footnotetext[*]{These authors contributed equally to this work.}

\begin{abstract}
Time-varying gravity field survey is one of the important methods for seismic risk assessment. To obtain accurate time-varying gravity data, it is essential to establish a gravity reference, which can be achieved using absolute gravimeters. Atom gravimeters, as a recently emerging type of absolute gravimeter, have not yet been practically validated for their reliability in mobile gravity surveys. To study and evaluate the operational status and performance metrics of the A-Grav atom gravimeter under complex field conditions, the University of Science and Technology of China, Hefei National Laboratory, and the Anhui Earthquake Agency conducted a joint observation experiment using an atom gravimeter (A-Grav) and relative gravimeters (CG-6) within the North China Seismic Gravity Monitoring Network. The experiment yielded the following results: 1) The standard deviations for mobile observations of the atom gravimeter is $2.1~\mathrm{\mu Gal}$; 2) The mean differences in point values and segment differences between the atom gravimeter and the relative gravimeter at the same locations is $5.8\pm17.1~\mathrm{\mu Gal}$ and $4.4\pm11.0~\mathrm{\mu Gal}$, respectively, with point value differences of less than $2.0~\mathrm{\mu Gal}$ compared to the FG5X absolute gravimeter at the same location; 3) The results of hybrid gravity adjustment based on absolute gravity control and the point value precision at each measurement point, with an average point value precision of $3.6~\mathrm{\mu Gal}$. The results indicate that the A-Grav atom gravimeter has observation accuracy and precision comparable to the FG5X absolute gravimeter, demonstrating good stability and reliability in field mobile measurements, and can meet the requirements for seismic gravity monitoring. This work provides a technical reference for the practical application of atom gravimeters in control measurements and time-varying gravity monitoring for earthquakes.

\end{abstract}
\maketitle

\section{Introduction}
Seismic gravity measurement is a crucial method for earthquake monitoring and research\cite{LiHui2009,scy2020,zyq2020}. Its primary task is to monitor the Non-terrestrial interior changes in the gravity field over time in tectonically active regions, and to study the temporal, spatial, and intensity variations of the gravity field during the processes of earthquake preparation, occurrence, and adjustment, as well as their underlying causes. To this end, China has gradually established a seismic gravity monitoring network covering the entire mainland, consisting of 105 absolute gravity points, 4,000 relative gravity points, and 86 continuous gravity stations\cite{scy2020}. Currently, a hybrid gravity measurement model combining absolute gravity observations and relative gravity surveys\cite{yyx2001} is employed to conduct repeated observations of the seismic gravity monitoring network. 
%This approach captures changes in the gravity field, serving earthquake monitoring, prediction, and related scientific research and national infrastructure development.
This methodology enables precise detection of gravitational field variations, thereby providing critical support for seismic hazard assessment (including earthquake monitoring and prediction), geophysical research, and the safeguarding of national infrastructure systems. 
Significant achievements have been made in this field\cite{CHENYUNTAI1979330,RUIHAO1983159,LiHui2009,CHONGYANG20121,zyq2010,zyq2013,zyq2017,chen2016,hu2021,JIA2023229676}.

High-precision absolute gravity observations play a significant role in monitoring changes in the gravity field:  

1) Conducting quasi-synchronous absolute gravity observations and relative gravity surveys provides a control benchmark for relative gravity data, enabling the acquisition of reliable time-varying gravity field information. This serves as a foundation for research on earthquake and disaster prediction\cite{LiHui2009}. 

2) Long-term absolute gravity observations provide precise data for studying earthquakes and crustal deformation processes\cite{ta2001,wy2004,sun2009,LELIN201161,QU201822,Wang2023}.  

During the past years, the absolute gravimeters used for seismic gravity measurements in China are the FG5 and A10 models, both developed and produced by the Micro-g LaCoste from U.S., the nominal accuracy is $2.0~\mathrm{\mu Gal}$ and $10.0~\mathrm{\mu Gal}$, respectively. The FG5X is the latest model deriving from the FG5 series. However, given the annual observation tasks involving thousands of points in the seismic gravity monitoring network, the limited number of available absolute gravity instruments hinders the full utilization of data from mobile ground gravity surveys and continuous gravity station measurements\cite{zyq2020}. Additionally, due to mechanical wear in traditional laser interferometers, their design life is limited, and instrument maintenance poses significant challenges\cite{Hu2012}. 

In recent years, absolute gravimeters(AG) based on cold atom interferometry have rapidly developed. Their measurement sensitivity and accuracy\cite{xie,li2023} are comparable to those of the classical optical interferometry absolute gravimeter FG5X. Complementing its high-performance capabilities, the AG has unique advantages such as faster measurement speeds, the ability to achieve long-term continuous measurements, no mechanical wear, and low maintenance costs\cite{cheng2022,m2018,fu2019,wu2019,li2023}. 
Given these characteristics, atom gravimeters are an attractive option for mobile absolute gravity measurements, provided they can be made compact and portable.
However, there is a lack of reports that have quantitatively evaluated the accuracy of atom gravimeter in real field environments. If AG is to be truly useful in mobile absolute gravity measurement scenarios, its mobile gravity surveys reliability must be verified.
%Therefore, the independent development of high-precision mobile absolute gravimeters and the corresponding performance verification studies under complex field conditions are of great practical significance for advancing the application capabilities of domestically produced absolute gravimeters.

%With advancements in the miniaturization, mobility, and engineering of atom gravimeters, several institutions both domestically and internationally have introduced mobile atom gravimeters. Currently, multiple institutions have conducted preliminary field observation experiments\cite{m2018,fu2019,wu2019,li2023}, laying the groundwork for the practical application of atom gravimeters in engineering. However, most of these studies were conducted under laboratory conditions or limited to localized areas with a focus on static observations. There is limited research on the operational performance of atom gravimeters under complex field conditions\cite{tan2020}, resulting in an insufficient understanding of their performance in such environments.  

This work based on the North China Seismic Gravity Monitoring Network, designed and conducted relevant experiments. 
%To verify the long-term measurement stability and observation accuracy of the A-Grav atomic gravimeter in field gravity station environments, a continuous absolute gravity observation experiment lasting over one month was conducted at the Jiufeng Seismic Station in Wuhan. The results demonstrated that the A-Grav atomic gravimeter achieves excellent stability and observation accuracy in this environment, providing potential for studying regional geological structural changes and seismic gravity anomalies. 
To verify the stability and observation accuracy of the A-Grav atom gravimeter in mobile environments, a hybrid gravity observation experiment was conducted in the field in conjunction with a relative gravimeter (CG-6), covering large spatial scales and significant gravity gradients. The results showed that the A-Grav atom gravimeter also exhibits good stability in such environments, with observation accuracy meeting the requirements for seismic gravity monitoring.
Following the initial experiment, the Nanjing site was re-measured with an FG5X absolute gravimeter to fulfill the requirements of the Chinese Crustal Movement Observation Network measurement campaign. Capitalizing on this opportunity, a direct comparative analysis between the USTC-AG12 and FG5X instruments was conducted. The results revealed a discrepancy of less than $2.0~\mathrm{\mu µGal}$ between the two instruments at the Nanjing site, thereby demonstrating the reliability of AG-based gravity measurements.
%Additionally, to assess the accuracy of the A-Grav atom gravimeter's observations, the absolute gravity observation results at the Nanjing reference point were compared with those obtained using the FG5X. The comparison confirmed that the A-Grav atom gravimeter achieves high measurement accuracy in mobile surveys.  
\begin{table*}[!]
	\renewcommand{\arraystretch}{1.5}
	\centering
	\caption{Nominal performance parameters of gravimeters}
	\label{t1}
	\resizebox{\linewidth}{!}{ % Resize table to fit within text width
		\begin{tabular}{cccccccc}
			\toprule
			\hline
			\hline
			\textbf{Model} & \textbf{Sensor Type} & \textbf{Range (mGal)} & \textbf{Resolution (\textmu Gal)} & \textbf{Absolute Zero Drift (\textmu Gal·d$^{-1}$)} & \textbf{Standard Deviation (\textmu Gal)} \\
			\midrule
			\hline
			CG-6 & Quartz Spring & 8000 & 0.1 & <200 & 5 &  \\
			A-Grav & Cold Atom & Global & 1 \textmu Gal & / & <5  \\
			\bottomrule
			\hline
			\hline
		\end{tabular}
	}
\end{table*}

This experimental study demonstrates that atom gravimeters significantly outperform conventional campaign-based gravity surveys by establishing a denser network of absolute gravity reference stations, thereby improving measurement traceability. The findings establish a methodological framework for deploying AG in both static absolute gravity control networks and dynamic seismic gravity monitoring systems, offering pivotal insights to accelerate the technological maturation and field operationalization of quantum sensor-based gravimetry.

\section{Gravity Survey}
\subsection{\label{sec:level2}Overview of the Experiment}
This work based on the North China Seismic Gravity Monitoring Network, designed and conducted a experiment to evaluate the performance of the A-Grav atom gravimeter under field conditions.  

%To verify the stability and measurement accuracy of the A-Grav atomic gravimeter in a field gravity station environment, a long-term continuous absolute gravity observation experiment was designed. This experiment was conducted from October 29, 2021, to December 3, 2021, in a cave at the Jiufeng Seismic Station in Wuhan. The A-Grav atom gravimeter was deployed for continuous absolute gravity observations over an extended period to assess its long-term stability and precision under real-world conditions. The results aimed to demonstrate the instrument's suitability for monitoring regional geological changes and seismic gravity anomalies.  

To further evaluate the stability, observation accuracy, and measurement precision of the A-Grav atom gravimeter in mobile field environments, a hybrid gravity observation experiment combining absolute gravity observations with relative gravity surveys and vertical gradient measurements was designed. This experiment was carried out in two phases.  

The first phase of observations took place from September 30, 2022, to November 22, 2022, at three measurement points along the Jinzhai-Liyang seismic gravity line: Jinzhai, Anqing, and Jingxian. Quasi-synchronous absolute gravity observations, relative gravity surveys, and vertical gradient measurements were conducted at these points. After the observations, the noise components of the A-Grav atom gravimeter were optimized to improve its performance.  

The second phase of observations was conducted from April 16, 2023, to April 23, 2023, at five gravity points: Huangshan, Hefei, Bengbu, Nanjing, and Huaibei. Simultaneous absolute gravity observations, relative gravity surveys, and vertical gradient measurements were carried out at these locations.  

\begin{figure}[H]
	\centering{\includegraphics[width=\linewidth] {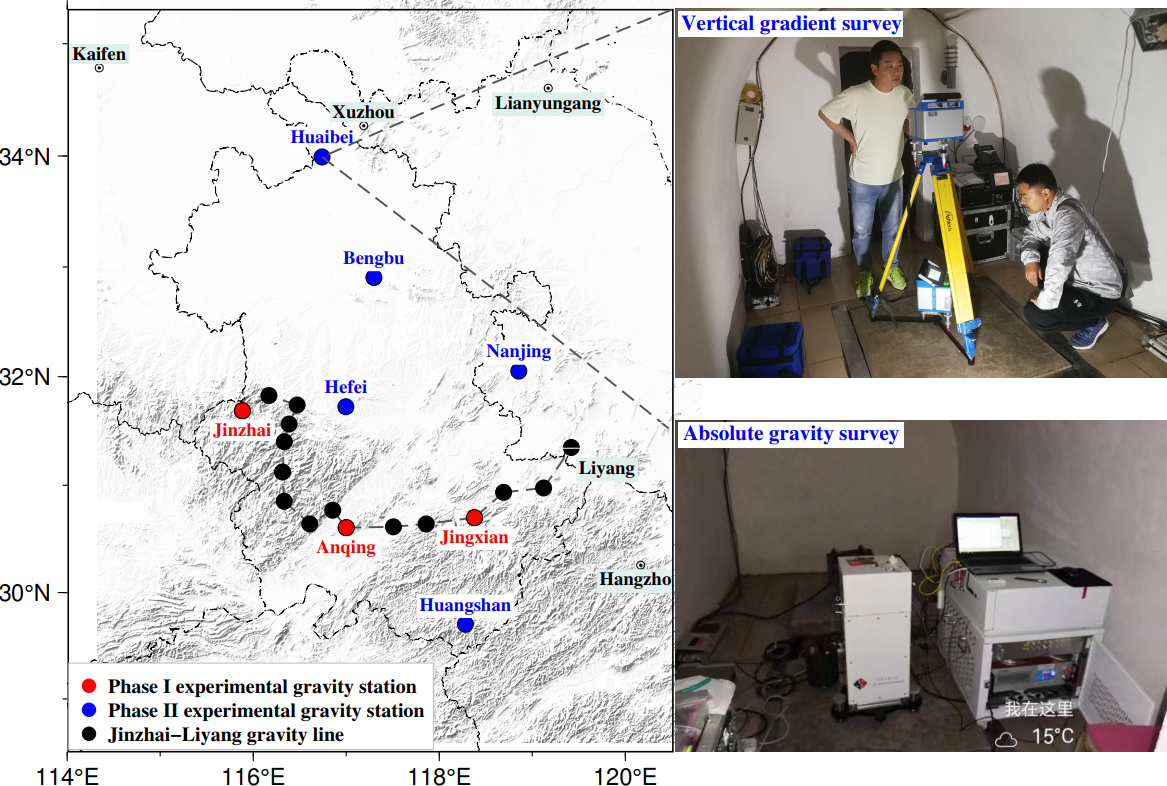}} 
	\caption{Gravity stations and lines of joint gravity survey experiment. The red points in the figure denote the survey stations from Phase I of the experiment, the blue points represent measurement locations surveyed in Phase II, and the black points form the Jinzhai-Liyang gravimetric profile.}
	\label{p1}
\end{figure}

The equipment used in these experiments included one A-Grav atom gravimeter\cite{li2023} and two CG-6(241/245) relative gravimeters. Key instrument parameters are listed in Table \ref{t1}, and the distribution of measurement points and lines is shown in Figure \ref{p1}.

\subsection{\label{sec:level3}Experimental Observations}
%\subsubsection{\label{sec:level4}Instrument Calibration}
%
%\subsubsection{\label{sec:level5}Vertical Gravity Gradient Measurement}

%\begin{table*}[!h] % 通栏表格使用 table* 环境
%	\centering
%	\caption{Gravity Observation Data}
%	\label{tab:gravity_observation}
%	\begin{tabular}{@{}ccccccc@{}} % @{} 去除表格两侧的空白
%		\toprule
%		No. & Site & Observation Instrument & Observation Date (y-m-d) & Gradient Segment Difference (\mu Gal) & Vertical Gradient (\mu Gal·cm^{-1}) & Height (cm) \\
%		\midrule
%		1. & Hefei &  & 2023-4-17 & 257.0208$\pm$0.7026 & 2.4021$\pm$0.0066 &  \\
%		2. & Huangshan &  & 2023-4-17 & 288.8732$\pm$0.7342 & 2.6997$\pm$0.0069 &  \\
%		3. & Bengbu &  & 2023-4-18 & 324.6523$\pm$0.3039 & 3.0341$\pm$0.0028 & 107 \\
%		4. & Nanjing &  & 2023-4-19 & 268.0552$\pm$0.5679 & 2.5052$\pm$0.0053 &  \\
%		5. & Huaibei & CG-6\ & 241/245 & 2023-4-20 & 245.1066$\pm$0.8322 & 2.2907$\pm$0.0078 &  \\
%		6. & Jinzhai &  & 2022-11-15 & 239.7360$\pm$0.3091 & 2.3504$\pm$0.0030 &  \\
%		7. & Anqing &  & 2022-11-16 & 291.3528$\pm$0.2950 & 2.8564$\pm$0.0029 & 102 \\
%		8. & Jingxian &  & 2022-11-19 & 228.2385$\pm$0.6351 & 2.2376$\pm$0.0062 &  \\
%		\bottomrule
%	\end{tabular}
%\end{table*}

\subsubsection{\label{sec:level6}Relative Gravity Survey}
The relative gravity survey was conducted using two CG-6(241/245) relative gravimeters in a synchronized manner. The survey employed a round-trip symmetric tandem observation mode, following the pattern A→B→A, to ensure high precision in the relative gravity measurements. Each survey line was closed within the shortest possible time on the same day to minimize errors caused by temporal variations in the gravity field.

The scale factor of an instrument is one of the critical factors affecting the accuracy of relative gravity measurements, directly influencing the accuracy of gravity difference measurements\cite{hao2016}. To avoid introducing additional systematic errors due to the time-varying characteristics of the instrument scale factor\cite{sun2002} and to ensure the accuracy and reliability of the observation results, the two CG-6(241/245) relative gravimeters were calibrated before each phase of the experimental observations.
The calibration results for both phases are shown in Appendix\ref{cali} Table \ref{t2}, with calibration accuracy better than $5\times10^{-5}$.

The vertical gravity gradient near the surface of the Earth is around $300~\mathrm{\mu Gal/m}$, there are significant differences in gravity values at different heights, therefore, a relative gravimeter is needed to make precise measurements and provide a gradient reference for the absolute value obtained by the atom gravimeter. The vertical gravity gradient observation method and results are presented in Appendix\ref{gg} and Table \ref{t3}. 
During this gravity surveys, the precision of the vertical gravity gradient segment differences was better than $\pm5.0~\mathrm{\mu Gal}$, and the precision of the vertical gravity gradients was better than $\pm0.025~\mathrm{\mu Gal\cdot cm^{-1}}$.
\begin{table*}[!]
	\renewcommand{\arraystretch}{1.5}
	\centering
	\caption{Phase I and II experiments result of relative gravity measurement.}
	\label{t6}
	\resizebox{\textwidth}{!}{
		\begin{tabular}{ccccccc}
			\toprule
			\hline
			\hline
			Survey Segment & Observation Date (y-m-d) & CG-6\#241(mGal) & CG-6\#245(mGal) & Mean Difference(mGal) & Mutual Difference Precision ($\mu$Gal) & Remarks \\
			\midrule
			Jinzhai-Jinzhai & 2022-11-15 & 0.2981 & 0.3027 & 0.3004 & 2.3 & \\
			Anqing-Anqing & 2022-11-16 & 5.8793 & 5.8836 & 5.8815 & 2.2 & Phase I observation\\
			Jingxian-Jingxian & 2022-11-19 & 0.2172 & 0.2096 & 0.2134 & 3.8 & \\
			\hline
			Huangshan-Hefei & 2023-4-17 & 197.8910 & 197.8900 & 197.8905 & 0.5 & \\
			Hefei-Bengbu & 2023-4-18 & 76.3070 & 76.3038 & 76.3054 & 1.6 & Phase II \\
			Nanjing-Bengbu & 2023-4-19 & 35.8415 & 35.8469 & 35.8442 & 2.7 & observation \\
			Bengbu-Huaibei & 2023-4-20 & 81.2340 & 81.2310 & 81.2325 & 1.5 & \\
			\hline
			\hline
			\bottomrule
		\end{tabular}
	}
\end{table*}

The measurement results for the 8 survey points using the relative gravimeter are shown in the table\ref{t6}.

\subsubsection{\label{sec:level7}Absolute Gravity Survey}
%The long-term continuous absolute gravity measurements and the absolute gravity measurements in the hybrid gravity surveys were conducted using the A-Grav cold atom absolute gravimeter. The long-term continuous absolute gravity observations were carried out at the Jiufeng Seismic Station in Wuhan, with a total observation duration of 36 days. 
During the hybrid gravity surveys, the absolute gravity measurements were performed synchronously with vertical gradient measurements and relative gravity surveys. After completing the vertical gradient measurements and relative gravity surveys at each measurement point, the atom absolute gravimeter was set up at the same location to conduct absolute gravity measurements. 

\begin{figure}[!]
	\centering{\includegraphics[width=\linewidth] {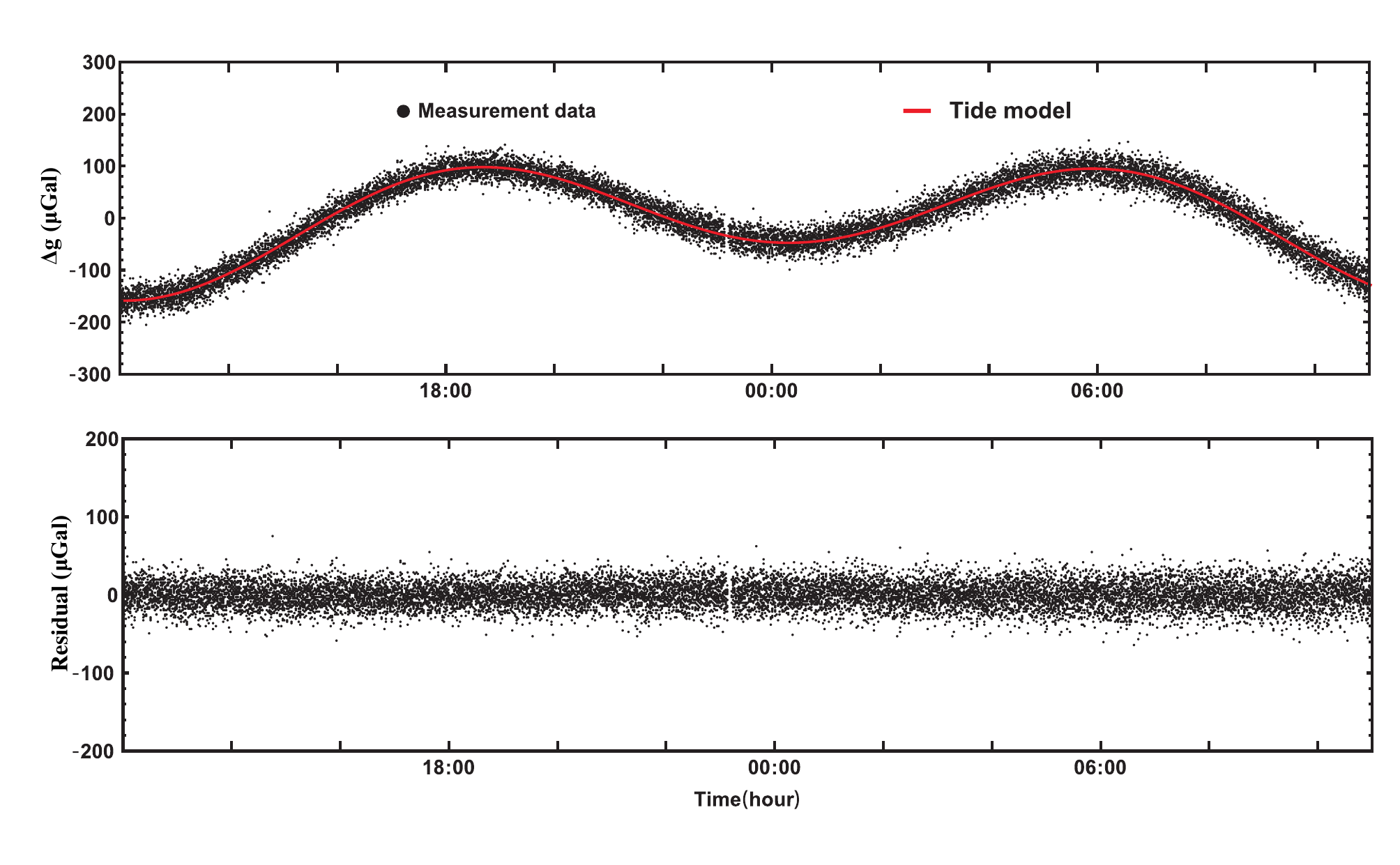}} 
	\caption{Raw gravity data from Nanjing point during the phase II. The black points represent gravity values measured at Nanjing station, whose fluctuations demonstrate tidal-induced gravitational variations, while the red curve shows theoretical values calculated using the Earth solid tide model. Each black data point was acquired at approximately 6-second intervals. The gravity residuals are obtained by subtracting the theoretical values from the measured gravity values.}
	\label{nanjing}
\end{figure}

Taking Nanjing as an example, there is a raw gravity data in figure\ref{nanjing} obtained by USTC-AG12. The red curve is the Earth solid tide model, which is calculate by $Tsoft$ software. The black points are the raw gravity value(all the values have been subtracted by their mean) measured by our atom gravimeter, and the residuals is subtracting the Earth solid tide, air-pressure and polar motion effect from the measured gravity. Similarly, the measurements at all other points in both phases of the experiment were processed using the same methodology. The observation duration ranging from 2.5 hours to 25.5 hours at each point with the sampling rate of 6 seconds.
%\begin{figure}[H]
%	\centering{\includegraphics[width=\linewidth] {jinzhai.pdf}} 
%	\caption{Raw gravity data from Jinzhai point during the phase I.}
%	\label{jinzhai}
%\end{figure} 

\begin{figure*}[!]
	\centering{\includegraphics[width=0.75\linewidth] {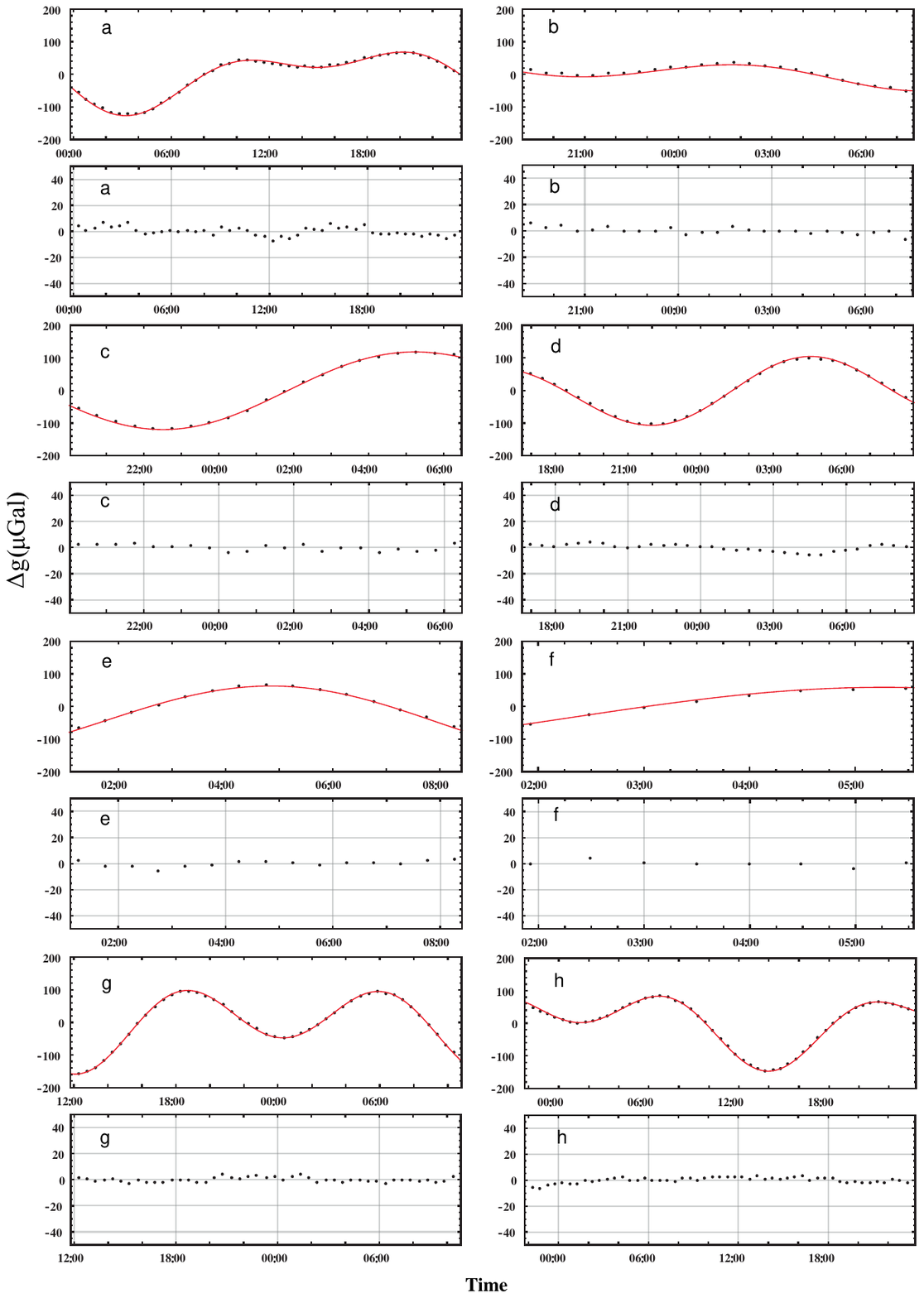}} 
	\caption{Distribution of absolute gravity observation and theoretical tidal model, gravity residuals in phase \uppercase\expandafter{\romannumeral1} and phase \uppercase\expandafter{\romannumeral2} experiments. (a), (b), and (c) represent the observation points of Jinzhai, Anqing, and Jingxian respectively from a single phase of observation; (d), (e), (f), (g), and (h) represent the observation points of Hefei, Huangshan, Bengbu, Nanjing, and Huaibei respectively from the second phase of observation.}
	\label{tr}
\end{figure*}

The measurement data from the atom gravimeter is recorded every 30 minutes, with the average of the valid absolute gravity data for that period serving as the observation value for that group. The observed values, tidal theoretical model values, and residual sequence distributions for each measurement point are presented, as shown in Figure \ref{tr}.
We use the standard deviation of the residual sequence, averaged over 30 minutes, as the criterion for assessing the validity of the data. A standard deviation of less than $5.0~\mathrm{\mu Gal}$ is considered acceptable, 
%as shown in figure \ref{p3}, 
the standard deviation results for all 8 points meet the requirements.
%\begin{figure}[!]
%	\centering{\includegraphics[width=\linewidth] {p3.pdf}} 
%	\caption{Between-group standard deviation of absolute gravity measurements in phase I and phase II experimental gravity stations.}
%	\label{p3}
%\end{figure}

At each location, Atom gravimeter needs to calibrate the systematic errors\cite{Peters,xht} related to the location, primarily the calibration of the tilt and the Coriolis force effect. In addition, the influence of the vertical gravity gradient should also be considered.
According to the vertical gradient results in Table \ref{t3}, the absolute gravity observation values at each measurement point are reduced to the height of the sensor of the relative gravity measurement instrument. Then, the reference datum is uniformly subtracted. Finally, the absolute gravity observation results at 8 measurement points are obtained, as shown in Table \ref{t5}. 
%For ease of comparison with relative gravimeters, a constant has been subtracted from each absolute gravity value in Table \ref{t5}. 
\begin{table*}[!]
	\renewcommand{\arraystretch}{1.5}
	\centering
	\caption{Results of Phase I and II Experiments of Absolute Gravity Measurement}
	\label{t5}
	\resizebox{\textwidth}{!}{
		\begin{tabular}{cccccccc}
			\toprule
			\hline
			\hline
			Point Name 	& \multicolumn{2}{c}{Observation Time} & Number of Observation Groups & Gravity Value at sensor height(mGal) & Standard Deviation between Groups($\mathrm{\mu Gal}$) & Weather & Remarks \\
			\midrule
			\hline
			Jinzhai & 2022 - 10 - 01/10 - 01 & 00:00 - 24:00 & 46 & 95.95384 & 3.5 & Sunny & \multirow{3}{*}{Phase I Observation} \\
			Anqing & 2022 - 11 - 17/11 - 18 & 19:00 - 08:00 & 25 & 53.58448 & 2.4 & Light Rain &  \\
			Jingxian & 2022 - 11 - 21/11 - 22 & 20:00 - 07:00 & 21 & 50.39148 & 2.3 & Light Rain &  \\
			\hline
			Hefei & 2023 - 04 - 16/04 - 17 & 16:40 - 09:00 & 32 & 144.03380 & 1.3 & Sunny &  \\
			Huangshan & 2023 - 04 - 18/04 - 18 & 01:00 - 09:00 & 15 & -53.85640 & 1.9 & Sunny & \multirow{4}{*}{Phase II Observation} \\
			Bengbu & 2023 - 04 - 19/04 - 19 & 01:30 - 05:30 & 8 & 220.33270 & 1.6 & Sunny &  \\
			Nanjing & 2023 - 04 - 20/04 - 21 & 12:00 - 11:00 & 45 & 184.48340 & 1.9 & Sunny &  \\
			Huaibei & 2023 - 04 - 22/04 - 23 & 22:00 - 24:00 & 51 & 301.58470 & 2.1 & Sunny &  \\
			\hline
			\hline
			\bottomrule
		\end{tabular}
	}
\end{table*}

\section{Experimental Results and Analysis}
Gravity adjustment constitutes a cornerstone data processing methodology in geophysics and geodesy, primarily employed to perform systematic corrections on gravity measurement datasets\cite{HWANG20021005}. This procedure serves three principal purposes: mitigating observational inaccuracies, enhancing data precision, and constructing a unified gravitational field model. At its core, this technique seeks to integrate discrete gravity observations into an internally consistent and coherent system with minimized residual errors through the application of sophisticated mathematical techniques.
The indirect adjustment model was employed to process the hybrid gravity observation data. This model uses the segment difference between two adjacent stations as the adjustment element. The basic observation equations are:
\begin{equation}
	\begin{aligned}
		& v_{ij} = \overline{g_i} - \overline{g_j} - \Delta g_{ij}\\
		& v_{ai} = \overline{g_{ai}} - g_{ai}
	\end{aligned}
\end{equation}
where \(\overline{g_i}\), \(\overline{g_j}\), and \(\Delta g_{ij}\) are the gravity estimates at points \(i\) and \(j\) and the segment difference after various corrections, respectively. \(v_{ij}\) is the adjustment element. \(\overline{g_{ai}}\) and \(v_{ai}\) are the estimate and residual of the absolute gravity observation \(g_{ai}\) at point \(i\). The gravity values at the observation points can be solved according to the least squares criterion. This work uses a weak baseline adjustment based on the above model, employing relative gravity surveys controlled by absolute gravity. The absolute gravity observations are weighted with a precision of \(5 \times 10^{-8} \, \text{m} \cdot \text{s}^{-2}\), and the weights of the relative gravity survey data are determined iteratively during the calculation process.

Using the observation results of the absolute gravity points in Table \ref{t5} as the control benchmark, combined with the synchronous relative gravity survey data in Table \ref{t6}, the differences and precisions between the adjusted gravity values and the observed gravity values at the Jinzhai, Anqing, and Jingxian points were obtained based on the above indirect adjustment model, as shown in Table \ref{t7}. The adjustment results for the four survey segments of Huangshan-Hefei, Hefei-Bengbu, Nanjing-Bengbu, and Bengbu-Huaibei are shown in Table \ref{t8}.
%\begin{figure}[!]
%	\centering{\includegraphics[width=\linewidth] {p5.pdf}} 
%	\caption{Difference between adjusted gravity values and observed values at measurement stations in phase I experiment.}
%	\label{p5}
%\end{figure}

%\begin{figure}[!]
%	\centering{\includegraphics[width=\linewidth] {p61.pdf}} 
%	\caption{Adjustment point value accuracy of gravity data of phase II experimental.}
%	\label{p6}
%\end{figure}

\begin{table*}[!]
	\renewcommand{\arraystretch}{1.5}
	\centering
	\caption{The difference between the adjusted gravity value and the observed value at the measurement stations of Jinzhai, Anqing, and Jingxian in phase I experiment.}
	\label{t7}
	\resizebox{\textwidth}{!}{
	\begin{tabular}{cccccc}
		\toprule
		\hline
		\hline
		Station Name & Observation Date (y-m-d) & Adjusted Gravity Value (mGal) & Point Precision ($\mu$Gal) & Adjusted Value Minus Observed Value ($\mu$Gal) & Mean Difference and Precision ($\mu$Gal) \\
		\midrule
		\hline
		Jinzhai & 2022-7-14 & 95.9705 & 7.6 & 16.7 & \\
		Anqing & 2022-7-14 & 53.5706 & 11.5 & -13.9 & $5.8 \pm 17.1$ \\
		Jingxian & 2022-8-5 & 50.4061 & 10.0 & 14.6 & \\
		\bottomrule
		\hline
		\hline
	\end{tabular}
	}
\end{table*}

\begin{table*}[!]
	\renewcommand{\arraystretch}{1.5}
	\centering
	\caption{Overall Adjustment Results of Gravity Data from the Phase II Experiment}
	\label{t8}
	\resizebox{\textwidth}{!}{
	\begin{tabular}{lcccccccc}
		\toprule
		\hline
		\hline
		Measurement Segment 
		& \multicolumn{2}{c}{CG - 6\#241} & \multicolumn{2}{c}{CG - 6\#245} & Mean Segment Difference & \multicolumn{2}{c}{Scale Value and Its Precision} \\
		& Segment Difference (mGal) & Self - difference ($\mu$Gal) & Segment Difference (mGal) & Self - difference ($\mu$Gal) & (mGal) & ($\mu$Gal) & CG - 6\#241 & CG - 6\#245 \\
		\midrule
		\hline
		Huangshan - Hefei 
		& 197.891 & 4 & 197.889 & 17 & 197.890  & \multicolumn{2}{c}{—} \\
		Hefei - Bengbu 
		& 76.304 & 0 & 76.301 & 4 & 76.303 & $1.000776\pm0.0$ & $1.000836\pm$ \\
		Nanjing - Bengbu 
		& 35.840 & 2 & 35.845 & 13 & 35.842 & 0.00034 & 0.000034 \\
		Bengbu - Huaibei 
		& 81.234 & 2 & 81.231 & 8 & 81.233 & \multicolumn{2}{c}{—} \\
		\hline
		\hline
		\bottomrule
	\end{tabular}
	}
\end{table*}

Based on Table \ref{t5}, for Phase I observations, the maximum and minimum intergroup standard deviations were 3.5 
$\mu$Gal and 2.3 $\mu$Gal, respectively, with a mean of 2.7 $\mu$Gal. For Phase II observations, the maximum and minimum intergroup standard deviations were 2.1 $\mu$Gal and 1.3 $\mu$Gal, respectively, with a mean of 1.8 $\mu$Gal. Both phases of observations were better than 5.0 $\mu$Gal. Compared to Phase I, the intergroup standard deviation in Phase II significantly decreased by 30\%, approximately $1.0~\mathrm{\mu Gal}$. As shown in Figure \ref{tr}, although the observation periods and duration of the A-Grav atom gravimeter varied at different measurement points, with the longest observation duration being 25.5 hours and the shortest being 2.5 hours, the observed values from the A-Grav atom gravimeter showed excellent consistency with the tidal values calculated by the tidal model during the observation period. The A-Grav atom gravimeter demonstrated good stability and adaptability in field mobile observations, with intergroup standard deviations better than 5.0 $\mu$Gal. It was able to accurately extract and identify tidal variation signals, and its performance has been further improved through technical iterations.

Table \ref{t7} shows the distribution of the differences between the adjusted gravity values at the measurement points in Jinzhai, Anqing, and Jingxian and the observed gravity values from the A-Grav atom gravimeter. The differences are the largest at the Jinzhai point, followed by the Jingxian point, and the smallest at the Anqing point, which are \(16.7\ \mathrm{\mu Gal}\), \(14.6\ \mathrm{\mu Gal}\), and \(- 13.9\ \mathrm{\mu Gal}\) respectively, with an average value of \((5.8\pm17.1)\ \mathrm{\mu Gal}\). 
The point precision at the Jinzhai point is the highest, which is \(7.6\ \mathrm{\mu Gal}\), followed by the Jingxian point with \(10.0\ \mathrm{\mu Gal}\), and the Anqing point has the largest value of \(11.5\ \mathrm{\mu Gal}\), and the average point precision values is \(9.7\ \mathrm{\mu Gal}\). The difference in gravity values between the same points is less than \(20\ \mathrm{\mu Gal}\), the measurements comply with the accuracy and precision specifications required for mobile gravity surveys.

As evidenced in Table \ref{t8}, for the CG-6(241/245) instruments, the maximum self-deviations are \(4.0\ \mathrm{\mu Gal}\) and \(17.0\ \mathrm{\mu Gal}\), the average self-deviations are \(2.0\ \mathrm{\mu Gal}\) and \(10.5\ \mathrm{\mu Gal}\), and the mutual differences are better than \(5.0\ \mathrm{\mu Gal}\). Meanwhile, the average point precision at the measurement points is \(3.6\ \mathrm{\mu Gal}\) during the phase II, which is better than \(5.0\ \mathrm{\mu Gal}\), meeting the precision requirements for seismic gravity observations. 
The calibration factors of CG-6(241/245) obtained from the adjusted calculation are \(1.000776\pm0.000034\) and \(1.000836\pm0.000034\), which are slightly different from the calibration factors of CG-6(241/245) in Table \ref{t2}, which are \(1.000751\pm0.000044\) and \(1.000798\pm0.000044\). The difference amplitudes are \(2.5\times10^{-5}\) and \(3.8\times10^{-5}\), which are less than \(5.0\times10^{-5}\). The calibration factor of the instrument directly affects the accuracy of the measured segment difference. According to the results of the relative gravity joint measurement in Table \ref{t6}, this paper gives the differences in the segment differences of the calibration factors of the CG-6(241/245) instruments in the four measurement segments of Huangshan-Hefei, Hefei-Bengbu, Nanjing-Bengbu, and Bengbu-Huaibei. The results show that the maximum influence in the Huangshan-Hefei measurement segment is \(4.9\ \mathrm{\mu Gal}\) and \(7.5\ \mathrm{\mu Gal}\), and the influences in other measurement segments are less than \(2.0\ \mathrm{\mu Gal}\) and \(3.1\ \mathrm{\mu Gal}\), with an average value of \(2.4\ \mathrm{\mu Gal}\) and \(3.7\ \mathrm{\mu Gal}\).
%\begin{figure}[!]
%	\centering{\includegraphics[width=\linewidth] {p9.pdf}} 
%	\caption{The difference of segment difference of the same segment of survey between absolute gravity and relative gravity observation in phase II experiment.}
%	\label{p9}
%\end{figure}
Table \ref{overall} shows the differences in the segment differences of the same measurement segments between the absolute gravity observations and the relative gravity joint measurements for the segments of Huangshan-Hefei, Hefei-Bengbu, Bengbu-Nanjing, and Huaibei-Bengbu. As evidenced in Table \ref{t6}, it can be known that the segment differences of the four measurement segments are relatively large, with an average value of 
\(97.8182\ \mathrm{mGal}\). Among them, the segment difference of the Huangshan-Hefei measurement segment is the largest, which is \(197.8905\ \mathrm{mGal}\), and the segment difference of the Bengbu-Nanjing measurement segment is the smallest, which is \(35.8442\ \mathrm{mGal}\).

%For the CG-6(241/245) instruments, the mutual difference precision ranges from the lowest of \(2.7\ \mathrm{\mu Gal}\) to the highest of \(0.5\ \mathrm{\mu Gal}\), with an average value of \(1.6\ \mathrm{\mu Gal}\), which is better than \(5.0\ \mathrm{\mu Gal}\). 
The differences in the segment differences of the same measurement segments are between \(6.5\ \mathrm{\mu Gal}\) and \(-19.5\ \mathrm{\mu Gal}\). Except that the difference in the Huaibei-Bengbu measurement segment is relatively large, which is \(-19.5\ \mathrm{\mu Gal}\), the differences in the segment differences of other measurement segments are all less than \(6.5\ \mathrm{\mu Gal}\), with an average value of \((4.4\pm11.0)\ \mathrm{\mu Gal}\), which is better than \(10.0\ \mathrm{\mu Gal}\). There is no obvious correlation between the segment differences and the differences in the segment differences.
\begin{table*}[!]
	\renewcommand{\arraystretch}{1.5}
	\centering
	\caption{The difference between the gravity segment difference value and the observed value at the phase II experiment.}
	\label{overall}
	\resizebox{\textwidth}{!}{
		\begin{tabular}{cccccccc}
			\toprule
			\hline
			\hline
			Measurement segment & gravity segment difference(mGal) & Observed gravity(mGal) & gravity difference($\mathrm{\mu Gal}$) & Mean Difference and Precision($\mathrm{\mu Gal}$)  \\
			\midrule
			\hline
%			Jinzhai & 95.9705 & 95.95384 & 16.7 &  & \multirow{3}{*}{Phase I Observation} \\
%			Anqing & 53.5706 & 53.58448 & -13.9 & $5.8~\pm17.1\mathrm{\mu Gal}$ &  \\
%			Jingxian & 50.4061 & 50.39148 & 14.6 &  &  \\
%			\hline
			Huangshan-Hefei & 197.8905 & 197.8902 & 0.3 & \\
%			Huangshan & 15 & -53.85640 & 1.9 & Sunny &  \\
			Hefei-Bengbu & 76.3054 & 76.2989 & 6.5 & $4.4~\pm11.0\mathrm{\mu Gal}$ \\
			Nanjing-Bengbu & 35.8442 & 35.8493 & -5.1 &  \\
			Bengbu-Huaibei & 81.2325 & 81.2520 & -19.5 & \\
			\hline
			\hline
			\bottomrule
		\end{tabular}
	}
\end{table*}

The above results indicate that in the field mobile gravity observations, the observation results of the A-Grav atom gravimeter are in good agreement with the observation results of the commercial CG-6 relative gravimeter with a nominal observation accuracy of 
\(10.0\ \mathrm{\mu Gal}\). The atom gravimeter has good stability and observation accuracy in the field, and the observation results are reliable.

\section{The Comparison of Absolute Gravimeters}
\begin{figure}[!]
	\centering{\includegraphics[width=\linewidth] {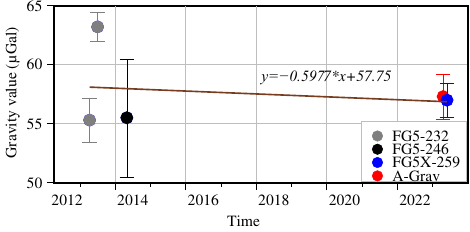}} 
	\caption{Long-term changes of absolute gravity observations at the Nanjing point. The geological structure of this point is relatively stable, and the gravity change does not exceed $10~\mathrm{\mu Gal}$ within a decade. The blue dots are the measurement results of FG5X-259, and the red dots are the measurement results of atom gravimeter A-Grav, The difference between them is better than $2~\mathrm{\mu Gal}$.}
	\label{p7}
\end{figure}

The most direct way to validate the accuracy of atom gravimeters is through synchronous comparative observations with the currently best commercially available laser interferometry absolute gravimeter, the FG5X. During the experimental observation phase, absolute gravity observations were conducted at the Nanjing point, with the two absolute gravity observations separated by no more than 40 days. The real-time observation data of A-Grav is shown in Figure \ref{nanjing}. Given the short interval between the two absolute gravity observations and the absence of significant seismic events or large-scale rainfall events in the Nanjing area during this period, the effects of seismic activity and local water load changes on gravity variations can be ignored. Additionally, combined with the long-term absolute gravity observation results at the Nanjing point shown in Figure \ref{p7}, this point is less affected by time-varying gravity fields and is stable over the long term, making the observation results at the Nanjing point suitable for studying the accuracy of atom gravimeters.

On May 30, 2023, the Hubei Earthquake Administration used the FG5X-259 absolute gravimeter to conduct 25 sets of observations at the Nanjing point, obtaining the absolute gravity value at a drop height of 130 cm, with an inter-group standard deviation of 1.4 $\mu$Gal. According to the vertical gradient of 2.5052 $\mu$Gal$\cdot$cm$^{-1}$ at the Nanjing point in Table \ref{t3}, the gravity value of the atom gravimeter at the Nanjing point in Table \ref{t5} was converted to the drop height of the FG5X-259 instrument. 

Following gradient-based reduction to a unified reference plane, the normalized measurements from both instruments demonstrated a good consistency, the gravity value of $g_{0}+184.2253 \mathrm{mGal}$ differed by only 0.3 $\mu$Gal from the observation result of the FG5X-259 gravimeter, $g_{0}+184.2250 \mathrm{mGal}$, where $g_{0}$ is a gravity value has been redacted in accordance with standard security protocols, this treatment will not affect the work of this paper.
the difference falls well within the acceptable threshold for high-precision gravity surveys. This marginal difference not only validates the technical maturity of our quantum-based solution, but also conclusively proves that our atom gravimeter has reached parity with conventional state-of-the-art devices in practical applications. Such comparable accuracy, achieved through fundamentally different operational principles, underscores our instrument's unique advantages in long-term stability, environmental adaptability, and potential for miniaturization.

%indicating that the accuracy of the A-grav atom gravimeter at the Nanjing point is in excellent agreement with the best commercially available optical interferometry absolute gravimeter, the FG5X-259. \textcolor{red}{re-write...}re-write...
%The nominal observation accuracy of the FG5X instrument is $2.0~\mathrm{\mu Gal}$, 
%and there is no significant systematic bias between different instruments of the same type.
%\begin{figure}[H]
%	\centering{\includegraphics[width=\linewidth] {p8.pdf}} 
%	\caption{Allan variance at Nanjing station.}
%	\label{p8}
%\end{figure}
%Figure \ref{p8} presents the Allan variance that characterizes the short-term stability (second-level stability) and long-term stability of the A-Grav atomic gravimeter. The red dots in the figure represent the calculated variances, and the red dashed line is the result of fitting using the ct method. The goodness of fit is greater than 0.99. It can be observed that at the Nanjing site, under the environmental conditions of field mobile measurement, the sensitivity (second-level stability) of the A-Grav atomic gravimeter can reach 
%\(37.2\ \mathrm{\mu Gal/\sqrt{Hz}}\), and the long-term stability is \(1.2\ \mathrm{\mu Gal\ @\ 10000\ s}\). This indicates that it exhibits good stability, observational accuracy, and precision in mobile seismic gravity measurements.

%\subsection{Analysis of mixed gravity measurement results}

\section{Conclusion and Outlook}
This work based on the North China Seismic Gravity Monitoring Network, combines the absolute gravimeter FG5X and the relative gravimeter CG-6 to investigate the working state of the A-Grav atom gravimeter in complex field environments through hybrid gravity measurements. The following conclusions are drawn:

\begin{enumerate}
%	\item Through long-term continuous absolute gravity observation experiments with the A-Grav atomic gravimeter at fixed points, it is verified that the mean intergroup standard deviation of the long-term continuous gravity residuals is $3.5 \pm 0.5 \, \mu\text{Gal}$, which is better than $5.0 \, \mu\text{Gal}$
%	. This indicates that the instrument has good stability and adaptability in long-term observation environments at stations and can accurately extract and identify tidal variation signals.
	\item Field hybrid gravity observation experiments show that during Phase I observations, the mean difference between the A-Grav and CG-6 observation point values is $5.8 \pm 17.1 \, \mu\text{Gal}$. During Phase II observations, the mean precision of the measurement points is $3.6 \, \mu\text{Gal}$, which is better than $5.0 \, \mu\text{Gal}$, and the mean difference in segment differences is $4.4 \pm 11.0 \, \mu\text{Gal}$. These results are in good agreement with the observations from the commercial CG-6 relative gravimeter, which has a nominal observation precision of $10.0 \, \mu\text{Gal}$. This indicates that the A-Grav atom gravimeter provides reliable results in mobile measurements, demonstrating good stability, adaptability, and observation precision.
	
	\item By comparing with the observation results from the absolute gravimeter FG5X, it is verified that the difference between the A-Grav atom gravimeter and the FG5X is less than $2.0 \, \mu\text{Gal}$. This indicates that the A-Grav atom gravimeter has good accuracy and observation precision in mobile measurements, meeting the requirements for seismic gravity monitoring.
\end{enumerate}

With the accumulation and enrichment of absolute gravity observation data, research on the relationship between absolute gravity changes and earthquakes has continued to develop. Wang Yong et al.\cite{wy2004} revealed significant gravity decreases of 14.8 $\mu$Gal and 10.9 $\mu$Gal at the Lijiang and Eryuan points, respectively, within a radius of 150 km from the epicenter, based on multi-period repeat observations using FG5 and JILAG absolute gravimeters before and after the 1996 Lijiang Ms7.0 earthquake. Xing Lelin et al.\cite{LELIN201161} discovered a significant upward gravity trend before the 2008 Wenchuan Ms8.0 earthquake, based on multi-period re-measurement data from the Chengdu Reference Station FG5 absolute gravimeter near the epicenter from 2002 to 2008, with a long-term cumulative change exceeding 20 $\mu$Gal and an annual change of 5.3 $\mu$Gal. Chen et al.\cite{chen2016} concluded that the Shigatse absolute gravity point, located 430 km west of the 2015 Nepal Mw7.8 earthquake, experienced a cumulative change exceeding 40 $\mu$Gal, with an annual rate of 22.4 $\mu$Gal, based on re-measurement data from four absolute gravity points on the Tibetan Plateau. Jia et al.\cite{JIA2023229676} revealed a synchronous upward gravity trend at four absolute gravity points within a radius of 350 km from the epicenter of the 2022 Menyuan Ms6.9 earthquake, based on multi-period re-measurement data from FG5 and A10 absolute gravimeters at four absolute gravity points on the southeastern edge of the Tibetan Plateau, with an annual change of 3.94 $\mu$Gal over the five years preceding the earthquake. These research results indicate that significant gravity field changes before strong earthquakes can be detected by the current high-precision commercial absolute gravimeter FG5X. The analysis results of hybrid gravity measurements show that atom gravimeters have observation accuracy and precision comparable to the FG5X absolute gravimeter, with intergroup standard deviations of long-term continuous observations better than $\pm$5.0 $\mu$Gal. These characteristics indicate that atom gravimeters can detect and extract gravity field changes that meet the requirements of seismic gravity monitoring. Moreover, they can obtain the complete time-varying evolution process of the gravity field before and after strong earthquakes through long-term continuous observations. This will provide valuable basic data for studying the temporal, spatial, and intensity variation characteristics and causes of the gravity field during the processes of earthquake preparation, occurrence, and adjustment, and will provide foundational research results for earthquake and disaster prediction studies.

\begin{acknowledgments}
We thank the Hubei Provincial Seismological Bureau for providing absolute gravity data from Nanjing station, and Dr. Li Zhong-ya for insightful discussions and exchanges.
We wish to acknowledge the support of the Innovation Program for Quantum Science and Technology, China (Grant No. 2021ZD0300601), the Center for Ocean Mega-Science, Chinese Academy of Sciences, the Senior User Project of RV KEXUE (Grant No. KEXUE2020GZ03), the National Natural Science Foundation of China (Grant No. 12025406), Anhui Initiative in Quantum Information Technologies (Grant No. AHY120000), and the Shanghai Municipal Science and Technology Major Project (Grant No. 2019SHZDZX01).

\end{acknowledgments}

\section*{Data Availability Statement}
The data that support the findings of this study are available from the corresponding authors upon reasonable request.

\appendix
\section{Calibration and Gradient Measurement of Relative Gravimeters}
\subsection{\label{cali}Instrument Calibration}   
Currently, short baseline calibration for relative gravimeters can be performed using two methods: based on gravity baseline fields or on measured data from the survey area. Both methods have been shown to yield equivalent results\cite{xu2023}.  
Phase I Observation: The scale factors of the instruments were calibrated using multi-period measured data from the North China Seismic Gravity Monitoring Network.  
Phase II Observation: The instruments were calibrated at the Jinzhai gravity short baseline field in the Dabie Mountains.
The calibration results shows in Table \ref{t2}.
 
\begin{table*}[htbp] % 通栏表格使用 table* 环境
	\renewcommand{\arraystretch}{1.5}
	\centering
	\caption{Instrument Calibration and Observation Data}
	\label{t2}
	\resizebox{\linewidth}{!}{
		\begin{tabular}{@{}cccccc@{}} % @{} 去除表格两侧的空白
			\toprule
			\hline
			\hline
			Observation Date & \multicolumn{2}{c}{Instrument Scale Value} & Segment Difference (mGal) & Calibration Site \\
			\cmidrule(lr){2-3}
			\hline
			& CG-6\#241 & CG-6\#245 & & \\
			\midrule
			January 12, 2022 & 1.000465$\pm$0.000041 & 1.000529$\pm$0.000041 & 448.8793 & Anhui Seismic Gravity Monitoring Network \\
			April 12, 2023 & 1.000751$\pm$0.000044 & 1.000798$\pm$0.000044 & 225.1605 & Dabie Mountain Jinzhai Gravity Baseline Field \\
			\hline
			\hline
			\bottomrule
		\end{tabular}
	}
\end{table*}

\subsection{\label{gg}Vertical Gravity Gradient Measurement}
The vertical gradient measurements at the observation points were conducted using two CG-6(241/245) relative gravimeters in a low-high-low or high-low-high round-trip measurement mode. This approach was designed to reduce the gravity observations obtained by the A-Grav atom gravimeter at the height of its measurement sensor to the desired specific position, ensuring alignment with the reference plane of the relative gravity survey.

Given the actual height of the atom absolute gravimeter's measurement sensor, the vertical gradient observation platform was set to a uniform height of $102~\mathrm{cm}$ during Phase I observations and $107~\mathrm{cm}$ during Phase II observations. At each measurement point, no fewer than three qualified independent observations were made per instrument. In total, vertical gradient measurements were completed at 8 measurement points, as shown in Table \ref{t3}.
\begin{table*}[!]
	\renewcommand{\arraystretch}{1.5}
	\centering
	\caption{Phase I and II experiments measurement result of vertical gradient on gravity stations}
	\label{t3}
	\resizebox{\textwidth}{!}{ % Resize table to fit within text width
		\begin{tabular}{ccccccc}
			\toprule
			\hline
			\hline
			\textbf{No.} & \textbf{Station} & \textbf{Instrument} & \textbf{Date (y-m-d)} & \textbf{Gradient Difference (\textmu Gal)} & \textbf{Vertical Gradient (\textmu Gal/cm)} & \textbf{Height (cm)} \\
			\midrule
			\hline
			1 & Hefei & & 2023-4-17 & 257.0208 ± 0.7026 & 2.4021 ± 0.0066 & \\
			2 & Huangshan & & 2023-4-17 & 288.8732 ± 0.7342 & 2.6997 ± 0.0069 & \\
			3 & Bengbu & & 2023-4-18 & 324.6523 ± 0.3039 & 3.0341 ± 0.0028 & 107 \\
			4 & Nanjing & & 2023-4-19 & 268.0552 ± 0.5679 & 2.5052 ± 0.0053 & \\
			5 & Huaibei & CG-6(241/245) & 2023-4-20 & 245.1066 ± 0.8322 & 2.2907 ± 0.0078 & \\
			\hline
			6 & Jinzhai & & 2022-11-15 & 239.7360 ± 0.3091 & 2.3504 ± 0.0030 & \\
			7 & Anqing & & 2022-11-16 & 291.3528 ± 0.2950 & 2.8564 ± 0.0029 & 102 \\
			8 & Jingxian & & 2022-11-19 & 228.2385 ± 0.6351 & 2.2376 ± 0.0062 & \\
			\bottomrule
			\hline
			\hline
		\end{tabular}
	}
\end{table*}

%\begin{table*}[!] % 通栏表格使用 table* 环境
%	\renewcommand{\arraystretch}{1.5}
%	\centering
%	\caption{Amplitude Factors of Tidal Waves}
%	\label{t4}
%	\resizebox{\textwidth}{!}{
%		\begin{ruledtabular}
%			\begin{tabular}{@{}cccccc@{}} % @{} 去除表格两侧的空白
%				\toprule
%				Left\footnote{\(\delta_0\) is the tidal factor determined from long-term observations of the superconducting gravimeter at Wuhan Station, data from reference\cite{xu2014}.}
%				Tidal Wave & Amplitude Factor \(\delta\) & Amplitude Factor \(\delta_0\) & Tidal Model \(\delta_1\) & Amplitude Difference \\
%				Symbol & (Cold Atom Gravimeter) & (Superconducting Gravimeter) & (DDW99) & \((\delta - \delta_1)/\delta_1\) \\
%				\midrule
%				\hline
%				Q1 & 1.12189$\pm$0.01187 & 1.18585 & 1.15418 & -5.39\% \\
%				O1 & 1.13914$\pm$0.00263 & 1.17966 & 1.15418 & -3.43\% \\
%				K1 & 1.10966$\pm$0.01160 & 1.15371 & 1.13531 & -3.82\% \\
%				N2 & 1.13027$\pm$0.01004 & 1.17977 & 1.16179 & -4.20\% \\
%				M2 & 1.12781$\pm$0.00256 & 1.17591 & 1.16179 & -4.09\% \\
%				K2 & 1.13948$\pm$0.11212 & 1.17198 & 1.16179 & -2.77\% \\
%				\bottomrule
%			\end{tabular}
%		\end{ruledtabular}
%	}
%\end{table*}

\nocite{*}
\bibliographystyle{unsrt}
\bibliography{aipsamp}% Produces the bibliography via BibTeX.

\end{document}